\documentclass[fleqn,usenatbib]{mnras}

\usepackage{newtxtext,newtxmath}
\usepackage{subfigure}
\usepackage{makecell}

\usepackage[T1]{fontenc}

\DeclareRobustCommand{\VAN}[3]{#2}
\let\VANthebibliography\thebibliography
\def\thebibliography{\DeclareRobustCommand{\VAN}[3]{##3}\VANthebibliography}

\usepackage{graphicx}	
\usepackage{amsmath}	

\usepackage{multirow}

\title[Morphology Classification]{From Images to Features: Unbiased Morphology Classification via Variational Auto-Encoders and Domain Adaptation}

\author[Xu et al.]{
Quanfeng Xu,$^{1,2,5}$\thanks{Contact Email: \href{mailto:xuquanfeng@shao.ac.cn}{xuquanfeng@shao.ac.cn}}
Shiyin Shen,$^{1,4}$\thanks{Contact Email: \href{mailto:ssy@shao.ac.cn}{ssy@shao.ac.cn}}
Rafael S. de Souza,$^{3}$\thanks{Contact Email: \href{mailto:rd23aag@herts.ac.uk}{rd23aag@herts.ac.uk}}
Mi Chen,$^{1,2}$
Renhao Ye,$^{1,2}$
Yumei She,$^{5}$
\newauthor
Zhu Chen,$^{6}$
Emille E. O. Ishida,$^{7}$
Alberto Krone-Martins$^{8,9}$
and Rupesh Durgesh$^{10}$
\\
$^{1}$Key Laboratory for Research in Galaxies and Cosmology, Shanghai Astronomical Observatory, Chinese Academy of Sciences, 80 Nandan Rd., Shanghai 200030, China \\
$^{2}$School of Astronomy and Space Sciences, University of Chinese Academy of Sciences, No. 19A Yuquan Road, Beijing 100049, People’s Republic of China \\
$^{3}$Centre for Astrophysics Research, University of Hertfordshire, College Lane, Hatfield, AL10~9AB, UK \\
$^{4}$Key Lab for Astrophysics, Shanghai 200234, China \\
$^{5}$School of Mathematics and Computer Science, Yunnan Minzu University, 2929 Yuehua Street, Kunming, 650500, China\\
$^{6}$Shanghai Normal University, 100 Guilin Road, Shanghai 200234, China \\
$^{7}$Université Clermont Auvergne, CNRS/IN2P3, LPC, F-63000 Clermont-Ferrand, France \\
$^{8}$Donald Bren School of Information and Computer Sciences, University of California, Irvine, CA 92697, USA \\
$^{9}$CENTRA/SIM, Faculdade de Ciências, Universidade de Lisboa, Ed. C8, Campo Grande, 1749-016, Lisboa, Portugal \\
$^{10}$Independent Researcher
}

\date{Accepted XXX. Received YYY; in original form ZZZ}

\pubyear{2023}

\begin{document}
\label{firstpage}
\pagerange{\pageref{firstpage}--\pageref{lastpage}}
\maketitle

\begin{abstract}
We present a novel approach for the dimensionality reduction of galaxy images by leveraging a combination of variational auto-encoders (VAE) and domain adaptation (DA). We demonstrate the effectiveness of this approach using a sample of low redshift galaxies with detailed morphological type labels from the Galaxy-Zoo DECaLS project. We show that 40-dimensional latent variables can effectively reproduce most morphological features in galaxy images. To further validate the effectiveness of our approach, we utilised a classical random forest (RF) classifier on the 40-dimensional latent variables to make detailed morphology feature classifications. This approach performs similarly to a direct neural network application on galaxy images. We further enhance our model by tuning the VAE network via DA using galaxies in the overlapping footprint of DECaLS and BASS+MzLS, enabling the unbiased application of our model to galaxy images in both surveys. We observed that DA led to even better morphological feature extraction and classification performance. Overall, this combination of VAE and DA can be applied to achieve image dimensionality reduction, defect image identification, and morphology classification in large optical surveys.

\end{abstract}




\section{Introduction}\label{sec:intro}

Galaxy images contain a wealth of information about their morphologies and structural properties. Traditionally, these morphologies were determined through visual inspection of the images, considering features such as bulges, disks, spiral arms, and bars. It has been shown that galaxy morphologies are closely linked to other physical properties such as color, stellar population, structural parameters, and even the environments in which they are found \citep[e.g.][]{1973ApJ...179..427S,1980ApJ...236..351D,Strateva_2001}. 

Deep learning techniques have made significant progress in computer vision classification of natural images in recent years, and they have also been applied to various morphology-related problems in astronomy. Astronomical images, a type of natural image with a low signal-to-noise ratio, can particularly benefit from these new developments in computer vision. For example, \citet{10.1093/mnras/stx2052} used convolutional neural networks (CNNs) to identify strong gravitational lenses in the Kilo Degree Survey. \citet{Aniyan_2017} applied CNNs to classify radio galaxies in archival data from the Very Large Array, and \citet{CIPRIJANOVIC2020100390} used CNNs to distinguish between merging and non-merging galaxies in simulated images at high redshift. \cite{RN10} utilized CNNs to classify galaxies in the Galaxy Zoo 2 catalogue \citep[GZ2;][]{Willett2013} into five morphological categories, achieving a high classification accuracy of 95.21\%. Other studies have also utilized deep learning techniques to study various astronomical phenomena such as identifying bars in galaxies \citep{cavanagh_bars_2020} and microlensing events \citep{mroz_identifying_2020}. 
\cite{vavilova2021machine} evaluated the effectiveness of an auto-classification algorithm on galaxy images in the Sloan Digital Sky Survey (SDSS) Ninth Data Release \citep{Ahn2012} and found that the accuracy of predicted galaxy morphology is largely independent of the redshift over the range $ 0.02$\textless z \textless$0.1$ \citep[see also][]{Huertas-Company_2018,10.1093/mnras/stab2142,10.1093/mnras/stac2697}.
Furthermore, many other algorithms based on supervised deep learning have been applied to various galaxy morphology-related problems, such as estimating morphological parameters \citep{2022ApJ...935..138G}, fitting light profiles \citep{Li_2022}, and image deblending \citep{Boucaud2020,2022PhRvD.106f3023W}.

Supervised deep learning algorithms are commonly used for morphological classification of galaxy images, but they require a large amount of expert effort to annotate the data. Moreover, galaxy images from different astronomical surveys may have significant variations in survey parameters, such as point spread function, magnitude, and surface brightness limits. As a result, a classification network trained on data from one survey may not be effective when applied to data from another survey.

Alternatively, unsupervised learning methods can discover internal features of galaxy images and are more adaptable to galaxy images under different survey parameters. For instance, \cite{10.1093/mnras/stz3006} succeeded in classifying images of astronomical targets into four morphological categories (spirals, S0/Sa galaxies, ellipticals, and stars) by extracting morphological features from images through growing neural gas and hierarchical clustering algorithms.
In another approach, \cite{hayat2021self} employed a self-supervised network to extract the morphological features of galaxy images in the GZ2 dataset \citep[see also][]{wei2022unsupervised,fielding2022classification,2018MNRAS.473.1108H}.

Both supervised and unsupervised learning algorithms for galaxy morphology typically begin with the extraction of high-dimensional features from raw image data, followed by dimensionality reduction to a lower-dimensional space for the final classification or clustering. Dimensionality reduction can also enhance the comprehensibility of the data by reducing noise in low SNR astronomical images \citep[see][]{Hovis_Afflerbach_2021}. In fact, various dimensionality reduction algorithms have been applied to different aspects of astronomical data \citep[e.g.][]{Kartaltepe_2015}.

In addition, as a low SNR image, the application of the dimensionality reduction algorithm on astronomical images may further reduce the noise and thus enhance the comprehensibility of the data \cite[see][]{Hovis_Afflerbach_2021}. Indeed, different dimensionality reduction algorithms have been applied to many aspects of astronomical data \citep[e.g.][]{Kartaltepe_2015}. For instance, \cite{10.1093/mnras/staa2651} applied principal component analysis (PCA) to images of galaxies in the Galaxy Zoo CANDELS GOODS-S sample \citep{Simmons2016} and found that the images of galaxies can be represented by just 12 eigengalaxies while retaining 96\% of their variance. However, as high-dimensional complex data, astronomical images may not be fully expanded in linear space, particularly for morphology-related data structure features. Therefore, deep learning methods for the nonlinear dimensionality reduction of galaxy images is worth exploring.

The Variational AutoEncoder (VAE) \citep{kingma2013auto} is a self-supervised learning method that combines an encoder and a decoder structure to map between images and features, ensuring that images with similar morphological features will also be similar in latent variable distribution. This has resulted in the VAE algorithm being applied to astronomy for feature extraction and image generation \citep{lanusse2021deep}. For instance, \cite{10.1093/mnras/stab734} used Vector Quantized VAE to extract morphological features of $r$-band images from SDSS DR7 data, and found that the latent variables were highly correlated with the physical properties of galaxies. Despite these advances, aspects of deep learning-based galaxy morphological feature extraction algorithms are still yet to be explored. One is how many dimensions of latent parameters can adequately express the basic morphological features of galaxies.

Recent studies have shown the potential of deep learning models in describing the morphology of galaxies through the use of latent variable parameter spaces. For example, \cite{hayat2021self} demonstrated that self-supervised learning could recover semantically useful representations of sky survey images, which can be used as features without needing labelled data. Using a 2048-dimensional latent variable parameter space, their method achieved similar accuracy as supervised models while using 2-4 times fewer labels for training. \citet{lanusse2021deep} simulated galaxy images from only 16-d latent variables and obtained a much better description of galaxy morphology than the traditional single S\'ersic model. \citet{Zhou_2022}  combined a convolutional autoencoder and a bagging-based  multiclustering model for morphological classification of galaxies in CANDELS.

Deep learning models also demonstrate robustness against naturally occurring perturbations in low signal-to-noise ratio astronomical images \citep{ciprijanovic2022deepadversaries}. \citet{ciprijanovic2023deepastrouda} showed that domain adaptation techniques can make the performance of deep learning models on morphology classifications consistent across different astronomical datasets. \citet{Vilalta_2019} present a new domain adaptation method that relies on comparable model complexity rather than source-target similarity, using active learning to reduce dependence on source data. \citet{10.1093/mnras/stab1677} apply domain adaptation techniques such as Maximum Mean Discrepancy (see more detail in Section \ref{sec:DA}), Domain-Adversarial Neural Network \citep[DANN;][]{ganin2016domain}, Fisher loss, and entropy minimization to train neural networks on simulation data, and then transfer these networks to telescope observations, thereby demonstrating significant improvements in classification accuracy. \citet{gilda2022unsupervised} utilizes unsupervised domain adaptation to accurately predict galaxies' star formation histories, by learning mappings between simulated galaxy models and present-day observations. Therefore, applying domain adaption algorithms to the dimensionality reduction tasks in astronomical images could further improve the ability to extract effective features from image data.

This paper is structured as follows. In Section \ref{sec:data} we describe the main galaxy sample in the DESI legacy survey and the GZ-DECaLS dataset, which offers the most detailed morphological classification to date. In Section \ref{sec:method} we introduce our use of the VAE model to perform non-linear dimensionality reduction of galaxy images. In Section \ref{sec:application}, we discuss our choice for the dimensionality of the latent space to perform morphological classification. 
Section \ref{sec:domian_adapt} discusses a domain adaption approach for galaxies outside the DECaLS footprint to make the dimensionality reduction algorithm robust against perturbations such as differences in point spread function, image resolution, and noise. Finally, we discuss our findings' results and implications in Section \ref{sec:conslusion}.

\section{Data}
\label{sec:data}



\subsection{Main galaxy sample in the DESI legacy survey}
\label{sec:desi}

The Main Galaxy Sample (MGS) is a sample of low redshift galaxies limited by flux and defined as having an r-band magnitude \textless$17.77$. Initially designed by the SDSS \citep{Strauss_2002}, it contains approximately 700,000 galaxies within the SDSS legacy survey footprint. Most galaxies in the MGS have obtained fiber spectroscopy, and their median redshift is about 0.1 \citep{Abazajian_2009}. This study uses images of the MGS from the DESI Legacy Imaging Surveys \cite[DESI,][]{Dey_2019}, which provide deeper images of the extragalactic sky in the $g$, $r$, and $z$ bands over an area of approximately 14,000 $deg^2$. The DESI Legacy Survey comprises three public projects: the Dark Energy Camera Legacy Survey (DECaLS), the Beijing-Arizona Sky Survey (BASS), and the Mayall z-band Legacy Survey (MzLS). These surveys use two different telescopes and complement each other regarding the sky coverage and the survey bands while differing slightly in their observational parameters, such as image resolution and exposure time. The parameters for each survey are listed in Table \ref{tab:desi}.

\begin{table}
	\centering
	\caption{Survey parameters of DESI Legacy imaging surveys.}
	\label{tab:desi}
    \setlength{\tabcolsep}{1.5mm}{
	\begin{tabular}{ccccccc} 
		\hline
        \multirow{2}*{Survey} & Telescope/ & \multirow{2}*{Bands} & \multirow{2}*{Footprint} & Pixel \\
         & Instrument & & & resolution \\
        \hline
        \multirow{2}*{DECaLS} & Blanco/ & \multirow{2}*{g, r, z} & NGC($\delta\leq$+32$^{\circ}$) & \multirow{2}*{0\farcs262} \\
         & DECam &  & +SGC & \\
        \hline
        \multirow{2}*{BASS} & Bok/ & \multirow{2}*{g, r} & \multirow{2}*{NGC ($\delta\geq$+32$^{\circ}$)} & \multirow{2}*{0\farcs454} \\
         & 90prime &  &  &  \\
        \multirow{2}*{MzLS} & Mayall/ & \multirow{2}*{z} & \multirow{2}*{NGC ($\delta\geq$+32$^{\circ}$)} &  \multirow{2}*{0\farcs262} \\
         & Mosaic-3 &  &  & \\
        \hline
	\end{tabular}}
\end{table}

\subsection{Galaxy Zoo DECaLS}
\label{sec:zoodecal}

\begin{figure}
	\includegraphics[width=\columnwidth]{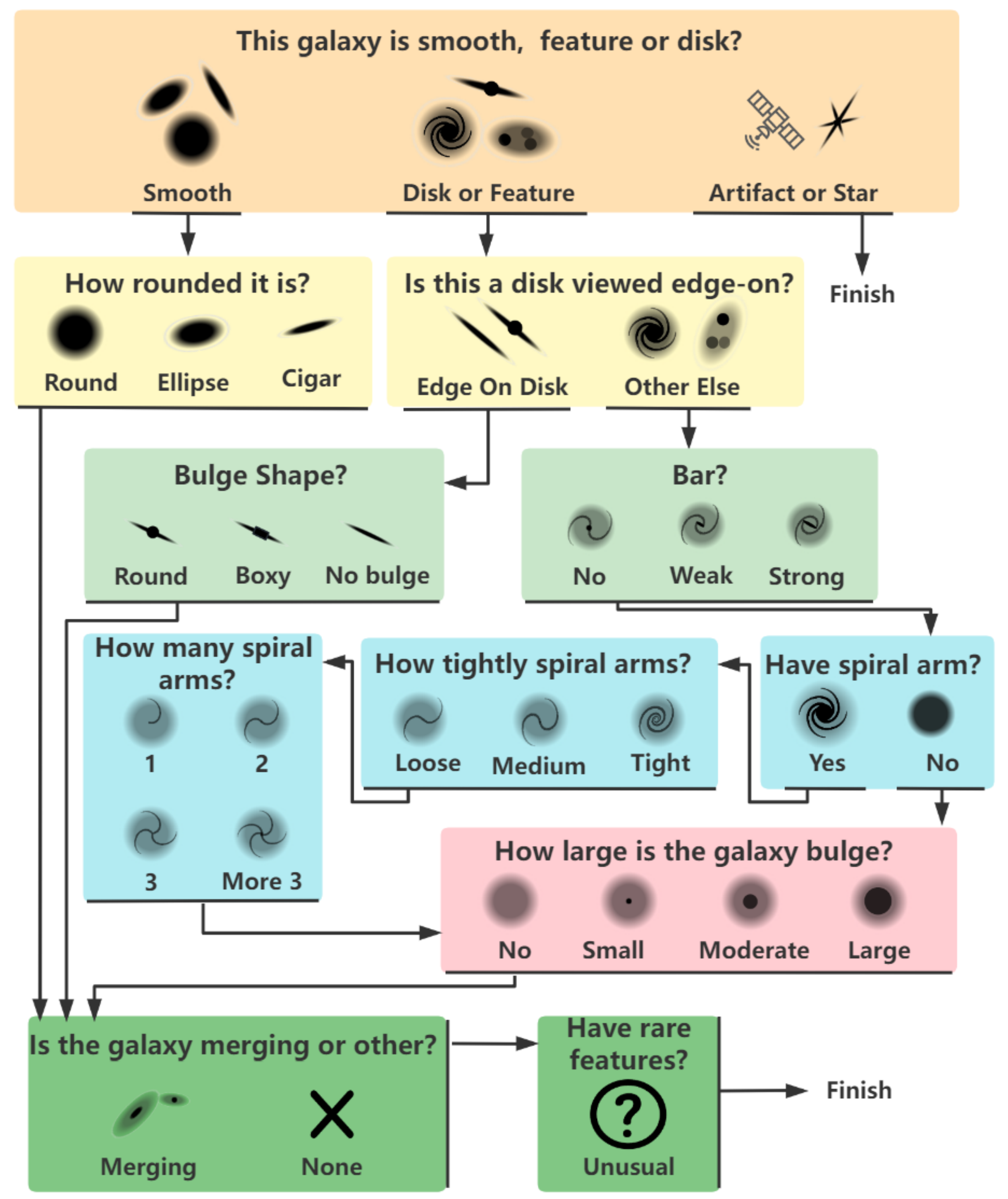}
    \caption{GZD-5 decision tree structure diagram. \citep[Adapted from ][]{Walmsley2022}}
    \label{fig:gzd}
\end{figure}

In the DECaLS footprint of the MGS, the Galaxy Zoo DECaLS project has provided visual morphology measurements for a sample of approximately 314,000 galaxies \citep{10.1093/mnras/stt1458}. Visual classifications of the volunteers for GZ DECaLS were collected during three campaigns, GZD-1, GZD-2, and GZD-5. The classifications were based on different decision trees, and the classified data were released in DECaLS Data Releases 1, 2, and 5, respectively. GZD-5 used an improved decision tree aimed at better identification of mergers and weak bars compared to GZD-1/2.
Using the GZD-5 classifications, \citet{10.1093/mnras/stt1458} trained an ensemble of Bayesian convolutional neural networks on the sample galaxies with at least three votes and successfully predicted the posteriors for the detailed morphology of all 314,000 galaxies under the GZD-5 decision tree. In the released catalog\footnote{\noindent \textbf{Galaxy Zoo Catalogues}: See \url{https://doi.org/10.5281/zenodo.4573248}}, each galaxy has been assigned likelihoods for each morphology label question following the GZD-5 decision tree (Fig. \ref{fig:gzd}).

In this study, we use the 314,000 GZ-DECaLS galaxies as the training sample of the VAE network to establish a mapping between the latent variables and morphological labels. For ease of use, we label the morphological parameters of each galaxy based on the probabilities of choices in each branch of the GZD-5 decision tree. Specifically, we split the GZD-5 decision tree into nine different morphological features: how rounded, edge-on, bulge shape, bar, have arm, arm tightness, arm count, bulge size, and merger. For the abovementioned 9 morphological features, we followed \citet{Walmsley2022} to arrange them in the decision tree and classified the presence or degree of each morphological feature by two to three to obtain a total of 24 characteristics. Using the decision tree and probability of morphological parameters, we identify 24 different morphological types for GZD-5 galaxies by setting the individual fraction threshold at 0.6. In order to keep the sample size as balanced as possible between the various types, we set the non-merger type threshold at 0.8. This was done to mitigate the impact of data imbalance on the model and results. The selection criteria and the resulting number of sample galaxies for each morphological type are listed in Table \ref{tab:morph}.

\begin{table}
\centering
\caption{Selection criteria and sample sizes for 24 galaxy samples and respective morphological parameters for a set of questions about their morphology. For each case, we provide the fraction threshold.}
	\label{tab:morph}
    \setlength{\tabcolsep}{1.8mm}{
	\begin{tabular}{ccccc} 
		\hline
        \multirow{2}*{Group} & \multirow{2}*{Question} & \multirow{2}*{Feature} & \multirow{2}*{Size} & Fraction \\
         &  &  &  & threshold \\
        \hline
        \multirow{3}*{Smooth} & \multirow{3}*{How round?} & Round & 54902 & \textgreater 0.6 \\
         & \multirow{2}* & Ellipse & 87929 &  \textgreater 0.6 \\
         &  & Cigar & 14502 &  \textgreater 0.6 \\
        \hline
        Disk or & \multirow{2}*{Edge on?} & Edge on & 9284 & \textgreater 0.6 \\
        Feature &  & Others & 49984 &  \textgreater 0.6 \\
        \hline
        \multirow{3}*{Edge-on} & \multirow{3}*{\makecell[c]{Bulge\\Shape}} & Round & 6314 &  \textgreater 0.6 \\
         &  & Boxy & 16 &  \textgreater 0.6 \\
         &  & No Bulge & 926 &  \textgreater 0.6 \\
        \hline
        \multirow{3}*{Not edge-on} & \multirow{3}*{Bar} & No Bar & 21087 &  \textgreater 0.6 \\
         &  & Weak Bar & 103 &  \textgreater 0.6 \\
         &  & Strong Bar & 1903 &  \textgreater 0.6 \\
        \hline
        \multirow{2}*{Not edge-on} & \multirow{2}*{Have arm?} & Arm & 44540 &  \textgreater 0.6 \\
         &  & No Arm & 1741 &  \textgreater 0.6 \\
        \hline
        \multirow{3}*{Have arm} & \multirow{3}*{\makecell[c]{Arm\\Tightness}} & Tight & 12702 &  \textgreater 0.6 \\
         &  & Medium & 315 &  \textgreater 0.6 \\
         &  & Loose & 2793 &  \textgreater 0.6 \\
        \hline
        \multirow{3}*{Have arm} & \multirow{3}*{\makecell[c]{Arm\\Count}} & 1 & 85 &  \textgreater 0.6 \\
         &  & 2 & 18103 &  \textgreater 0.6 \\
         &  & 3 & 113 &  \textgreater 0.6 \\
        \hline
        \multirow{3}*{Edge-on} & \multirow{3}*{\makecell[c]{Bulge\\Size}} & No Bulge & 484 &  \textgreater 0.6 \\
         &  & Small & 13239 &  \textgreater 0.6 \\
         &  & Moderate & 7973 &  \textgreater 0.6 \\
        \hline
        \multirow{2}*{All galaxies} & \multirow{2}*{Merger} & Merging & 4182 &  \textgreater 0.6 \\
         &  & No Merger & 86326 &  \textgreater 0.8 \\
        \hline
	\end{tabular}}
 \end{table}

\section{Method}
\label{sec:method}

\subsection{Variational Autoencoders}
\label{sec:vae}

The VAE architecture \citep{kingma2013auto} consists of an encoder and a decoder, each with its own set of learnable parameters. The encoder, with encoding parameters $\phi$, maps the high-dimensional input data $x$ to a low-dimensional latent variable representation $z$ through the probability distribution $p_\phi({z}|{x})$. With decoding parameters $\psi$, the decoder maps the latent variables back to the high-dimensional feature space, generating the reconstructed data $\hat{x}$ by $p_\psi(x|z)$ and $z$. Both $p_\phi({z}|{x})$ and  $p_\psi(x|z)$ are parameterized by a neural network.

In addition to the encoder and decoder networks, the VAE architecture includes a stochastic variational inference step. During this step, the model computes the mean $\mu(z)$ and variance $\sigma(z)$ of the latent variable features. These statistics are used to construct a high-dimensional normal distribution for all latent variables. The model then samples the latent variable features from this distribution to generate the corresponding high-dimensional data.

By using the stochastic variational inference technique, the VAE can learn a probability distribution over the latent variables that capture the underlying structure of the input data. The distribution is typically assumed to be a multivariate Gaussian with diagonal covariance, allowing efficient computation and sampling
\begin{equation}
    \mathbf{z} \sim q_\phi(z) = \mathcal{N}\left (\mu(z),\mathrm{diag}(e^{\ln\sigma(z)^2})\right)\, ,\mathbf{z} \in {\mathbb R^{N_L}}.
\end{equation}
With $q_\phi(z)$, the reconstructed data then is simply $\hat{x} \in q(\hat{x}) = p_\psi(x|z)q_\phi(z)$. 
VAE employs \textit{Kullback–Leibler} distance \citep[KL;][a measure of relative entropy]{Kullback1951} to map the difference between the distribution of $q_\phi(z)$ and $p(z)$, where $p(z)$ is the distribution of $z$ output from the encoder, i.e. $p_\phi(z|x)p(x)$. As a result, the global loss function of the VAE network can be written as
\begin{equation}\label{eq4}
    \mathcal{L}_\text{VAE} = \mathcal{L}_\text{rec} + k \cdot \mathcal{L}_\text{KL} = ||x-\hat{x}|| + k \cdot D_{\mathrm{KL}} (q_\phi(z)||p(z)).
\end{equation}
Here $k$ is a hyper-parameter that balances two loss components. In this study, we set $k$ as 1. Besides $k$, the number of latent variables $N_L$ is the other hyper-parameter affecting the reconstruction effect in the VAE model, which we will probe in detail in Section \ref{sec:dimension}. 

Overall, the VAE architecture provides a robust framework for learning compressed representations of high-dimensional data that can be used for various downstream tasks such as image generation, data compression, and data analysis.

\begin{figure}
	\includegraphics[width=0.5\textwidth]{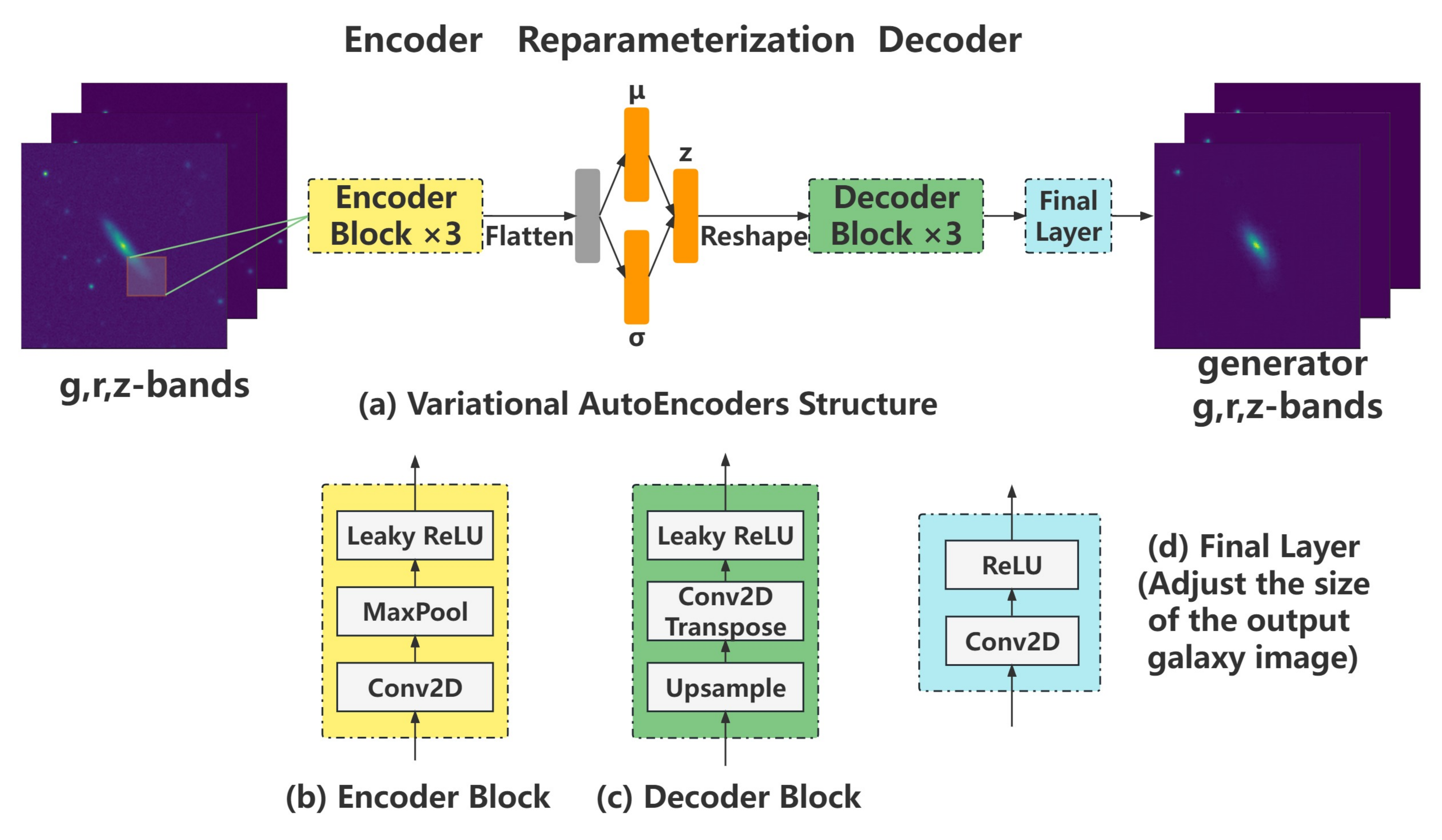}
    \caption{Structural details of the Variational Autoencoders in this study.}
    \label{fig:vae}
\end{figure}

The VAE model structure in this paper is illustrated in Figure \ref{fig:vae}, consisting of an encoder, a latent variable reconstruction, a decoder, and a final layer. The encoder comprises three encoder blocks, each containing a convolutional layer, a maximum pooling layer, and an activation layer. The three input channels of the first encoder block correspond to the $g$, $r$, and $z$ bands, and the subsequent blocks build upon the previous block's output. The three encoder blocks have 32, 64, and 128 output channels. The encoder's high-dimensional output vectors are then flattened into two $N_L$-dimensional latent variables, $\mu$ and $\sigma$, through two fully connected layers. For the decoding part, a $N_L$-dimensional latent variable feature $z$ is sampled from high-dimensional normal distributions with mean $\mu$ and scatter $\sigma$. The decoder comprises three decoder blocks, each containing a transposed convolution, an upsampling layer, and an activation function. The input channels for each decoder block are 128, 64, and 32, respectively. Finally, the convolutional layer and ReLU activation function of the final layer reconstruct images to their original sizes.

\subsection{Domain Adaption}
\label{sec:DA}

Deep learning models may struggle to perform well on different learning tasks due to differences in the data distributions between the source and target domains. This can be caused by diverse underlying astrophysical populations, but also, in the case of data coming from more than one observatory, differences in detector characteristics and observational conditions, to cite a few. In practical applications of deep learning, these differences in underlying statistical distributions can limit the generalization power of models. For example, feature extraction networks used for extracting galaxy morphological features through VAE may not be robust due to variations in data sources. 
Here we use domain adaptation to overcome some of these issues. Since images of the same celestial region obtained from different surveys intrinsically share identical physical properties, we sought to eliminate image differences caused by extrinsic variations in observational effects.

To quantify the difference between the source and target domains, \cite{pan2008transfer} proposed the maximum mean difference (MMD). MMD is defined as:

\begin{equation}
\label{eq7}
\mathrm{MMD} \left(\mathcal{F},X_S,X_T\right) = \sup_{f \in \mathcal{F}} \left(\mathop{\mathbf{E}}\limits_{x_s \in X_S}[f(x_s)] - \mathop{\mathbf{E}}\limits_{x_t \in X_T}[f(x_t)]\right).
\end{equation}
where $X_S$ and $X_T$ represent the source and target domain data, and $\mathcal{F}$ is a space of functions. The equation defines the supremum of the expectation of the functional mapping between the source and target domains. Homogeneous transfer learning reduces the difference in data distribution between the source and target domains by decreasing the MMD. We will discuss the usage of MMD in domain adaptation in Section ~\ref{sec:domainadapt}.

\section{Application to the MGS in DECaLS}
\label{sec:application}

In this section, we use the VAE model to reduce the dimensionality of the $g,r,z$ band images of the MGS in DECaLS survey. In Section ~\ref{sec:dimension}, we explore the optimised dimension of the latent variables in the VAE model. And in Section ~\ref{sec:Visualization}, we discuss the properties of the latent variable distribution and potential scientific applications. We then test the effectiveness of the morphology feature classification using the resulting latent variables from VAE model in Section ~\ref{sec:classification}.

Throughout the study, we use the Adam \citep{2014arXiv1412.6980K} optimizer for networks and set the learning rate to 0.0001, batch size to 512. We train this network for 100 epochs.


\subsection{The dimensionality of latent variables}
\label{sec:dimension}

The dimension of the latent variable features $N_L$ is the only hyper-parameter in our VAE model. In principle, a higher $N_L$ would, of course, contain more information, so that more comprehensive morphological features of galaxy images would be contained. On the other hand, a dimensionality reduction algorithm seeks to reduce the dimension of data as much as possible while preserving majority of the data feature. Thus, it is necessary to choose an appropriate dimension to balance the image reconstruction quality and the dimension of latent variables.

In this study, we tend to limit $N_L$ in the range of 10 to 100. If the dimension of latent variables is too low (e.g. \textless10), we do not expect that the morphological information of galaxy images could be adequately contained by the latent variables. On the other hand, if the dimension of latent variables is too high (e.g. \textgreater100), the dimensionality reduction would not be very significant, and the resulting interpretability of the latent parameter becomes weaker. Based on the above considerations, we test 6 cases with $N_L$ being 10, 20, 40, 60, 80, and 100 respectively.

We use the structural similarity index measure (SSIM) proposed by \cite{wang2004image}, which is a measure of the similarity between the reconstructed image and the original. 
In this algorithm, the luminance $l\left(x_{ij},\hat{x_{ij}}\right)$, contrast $c\left(x_{ij},\hat{x_{ij}}\right)$, and structure $s\left(x_{ij},\hat{x_{ij}}\right)$ of $N\times N$ sliding windows at the same position $(i,j)$ of two images are first compared,
\begin{equation}\label{eq5}
\begin{cases}
l\left(x_{ij},\hat{x_{ij}}\right)=\frac{2\mu_x\mu_{\hat{x}}+C_1}{\mu_x^2+\mu_{\hat{x}}^2+C_1},\\
c\left(x_{ij},\hat{x_{ij}}\right)=\frac{2\sigma_x\sigma_{\hat{x}}+C_2}{\sigma_x^2+\sigma_{\hat{x}}^2+C_2},\\
s\left(x_{ij},\hat{x_{ij}}\right)=\frac{\sigma_{x\hat{x}}+C_3}{\sigma_x\sigma_{\hat{x}}+C_3},
\end{cases}
\end{equation}
where $\mu_x,\mu_{\hat{x}}$ and $\sigma_x^2, \sigma_{\hat{x}}^2$ are the mean and variance of the original and the reconstructed image windows respectively, $\sigma_{x\hat{x}}$ is the covariance between the original and the reconstructed images in the same windows, and $C_1, C_2, C_3$ are three constants. Following \cite{wang2004image}, we set $C_1=(0.01L)^2, C_2=(0.03L)^2, C_3=C_2/2$, and $L$ is the dynamic range of pixel values (255 for 8-bit grayscale images). The SSIM function is defined as \begin{equation}\label{eq6}
\begin{aligned}
\mathrm{SSIM} \left(x_{ij},\hat{x_{ij}}\right) &= l\left(x_{ij},\hat{x_{ij}}\right)^\alpha \cdot c\left(x_{ij},\hat{x_{ij}}\right)^\beta \cdot s\left(x_{ij},\hat{x_{ij}}\right)^\gamma.
 \end{aligned}
\end{equation}
Here, we set $\alpha=\beta=\gamma=1$, following \cite{wang2004image}. The global SSIM parameter is then calculated through the average values of continuously sliding windows.

From equation \eqref{eq6}, the image similarity value evaluated by SSIM is between 0 and 1. The closer the similarity value is to 1, the more similar the images. 

\begin{figure}
	\includegraphics[width=\columnwidth]{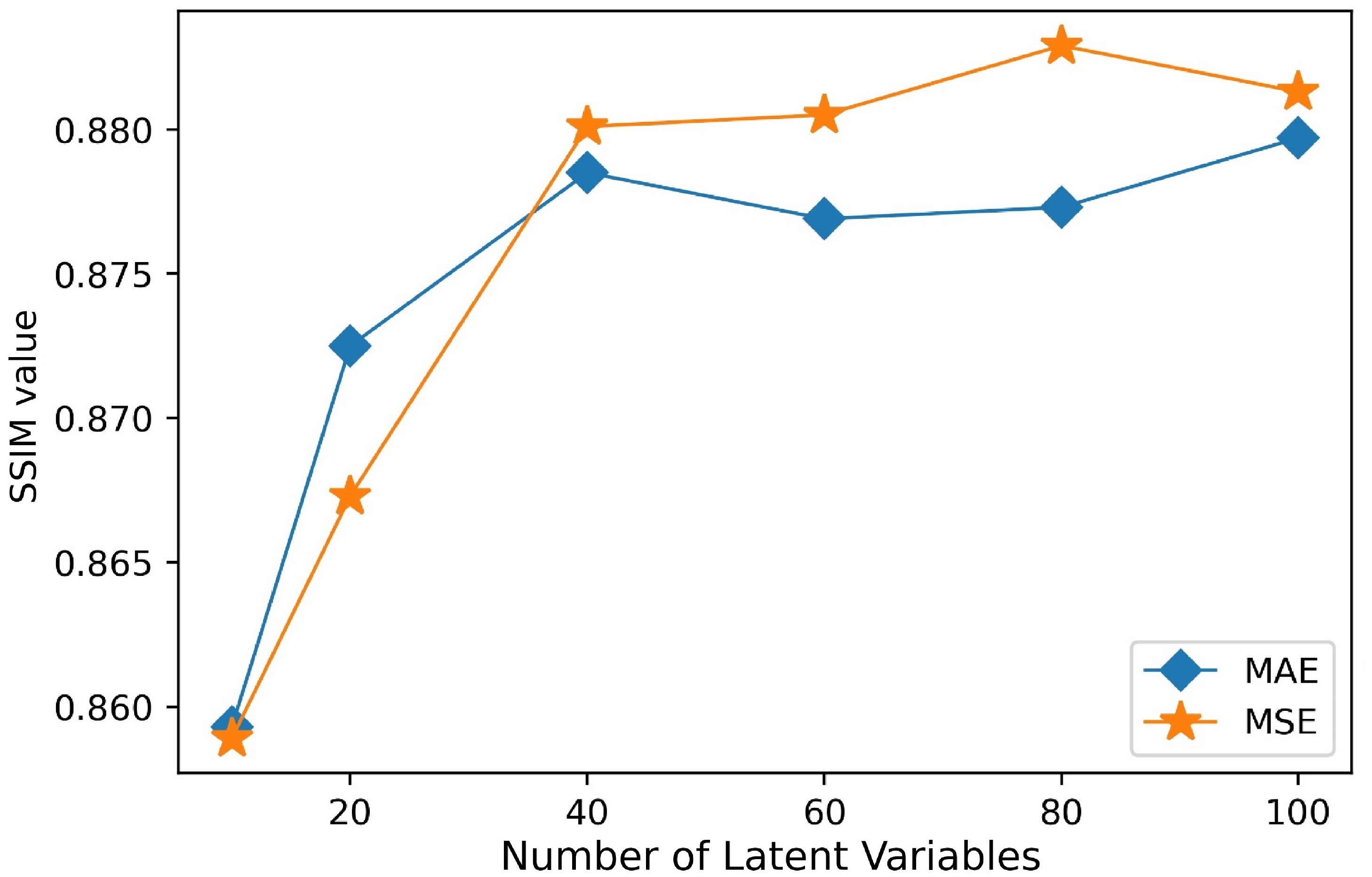}
    \caption{Evaluation of SSIM reconstructed values in two loss functions with different dimensionality of latent variables for all images.}
    \label{fig:ssim_loss}
\end{figure}

\begin{figure}
	\includegraphics[width=\columnwidth]{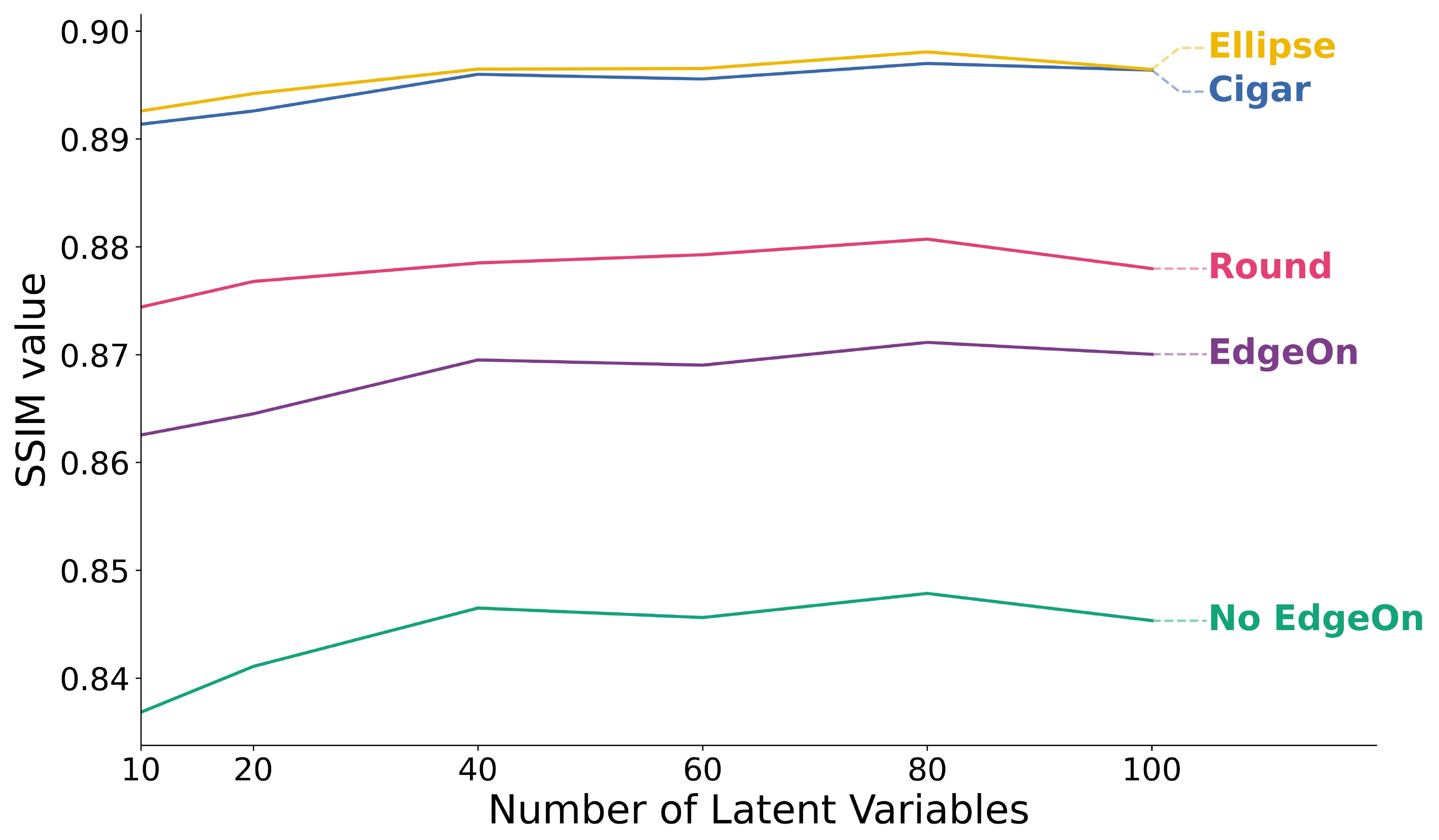}
    \caption{Evaluation of different latent variable dimensions in various categories of SSIM reconstructed values.}
    \label{fig:ssim_sum}
\end{figure}

We evaluate the reconstruction performance for different dimensionality of latent parameters by the SSIM values of the reconstructed and the original images. Moreover, we adopt two loss functions, Mean Absolute Error (MAE) and Mean Squared Error (MSE), to optimize the VAE model. Figure \ref{fig:ssim_loss} shows the reconstruction performance in different cases. As expected, SSIM values increase monotonously with $N_L$. Moreover, we find that SSIM values present a inflection point at $N_L\sim40$. 

In addition, we compare the reconfiguration performance for galaxies with different morphology features. For each type of restructuring evaluation, only the first stage branch of the decision tree (the classification of yellow background in Figure \ref{fig:gzd}) is selected for analysis and comparison. Figure \ref{fig:ssim_sum} shows that the VAE model performs better in reconstructing images of cigar-shaped galaxies and elliptical galaxies, while poorer in images of disk galaxies with detailed structures. However, the SSIM evaluation indexes for different categories all show inflection points around $N_L\sim 40$. Thus, considering the dimensionality of latent variables and the quality of reconstructed images, we select the experimental results with the loss function of MSE and latent variable parameters $N_L=40$ for further investigation.

\subsection{Visualization of Latent Variables}
\label{sec:Visualization}

\begin{figure}
    \centering
    \subfigure[Binary classification]{
        \includegraphics[width=0.4\columnwidth]{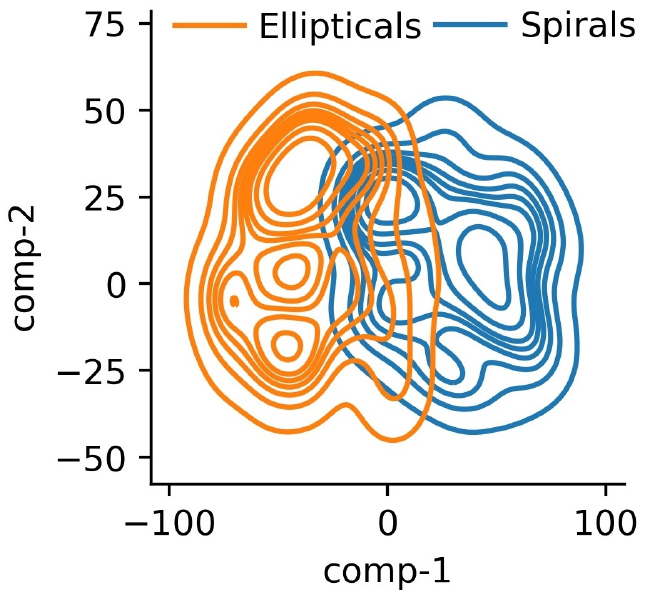}
        \label{fig:mult_class}
    }
    \subfigure[Multiclass classification]{
        \includegraphics[width=0.55\columnwidth]{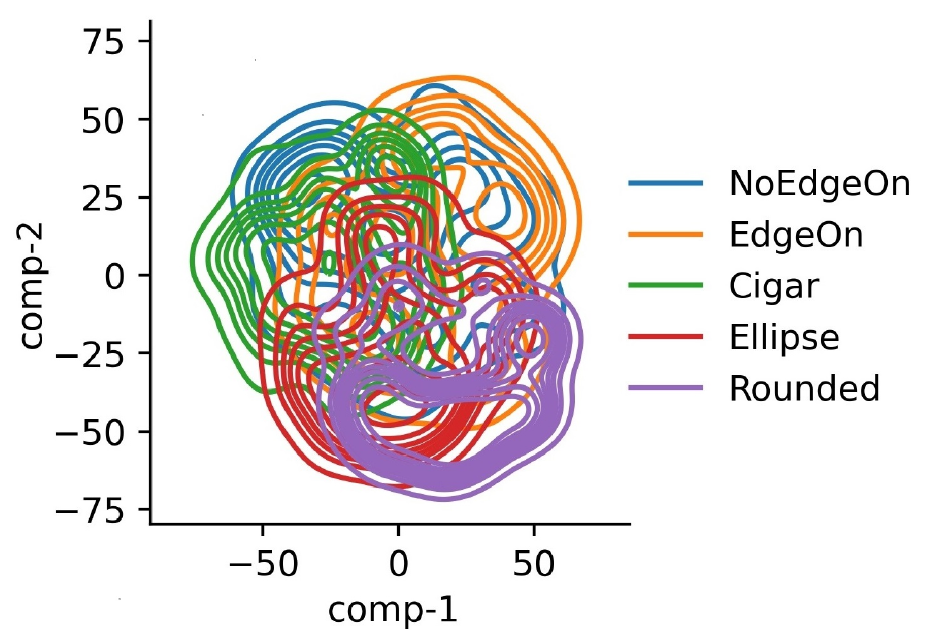}
        \label{fig:two_class}
    }%
    \caption{The plot displays t-SNE projections of our galaxy sample, with broad categorizations (Spirals vs. Ellipticals) and narrower classifications (No edge-on, edge-on, cigar, ellipse, rounded), each color-coded. As classifications refine, the projected latent space becomes increasingly intermingled.}
    \label{fig:tsne}
\end{figure}

When projected onto a latent space, similar morphological features tend to cluster. We employ t-SNE \citep{van2008visualizing} to reduce the 40-dimensional space to two dimensions, as illustrated in Figure \ref{fig:tsne}. This figure showcases the density contours of the latent space. The left panel presents broader categories, while the right offers finer details. As the granularity increases, the latent space becomes denser and more intertwined. Importantly, we remind that this is just a visual representation of a 40-dimensional space where we expect morphological features to be more distinct.

\begin{figure}
	\includegraphics[width=\columnwidth]{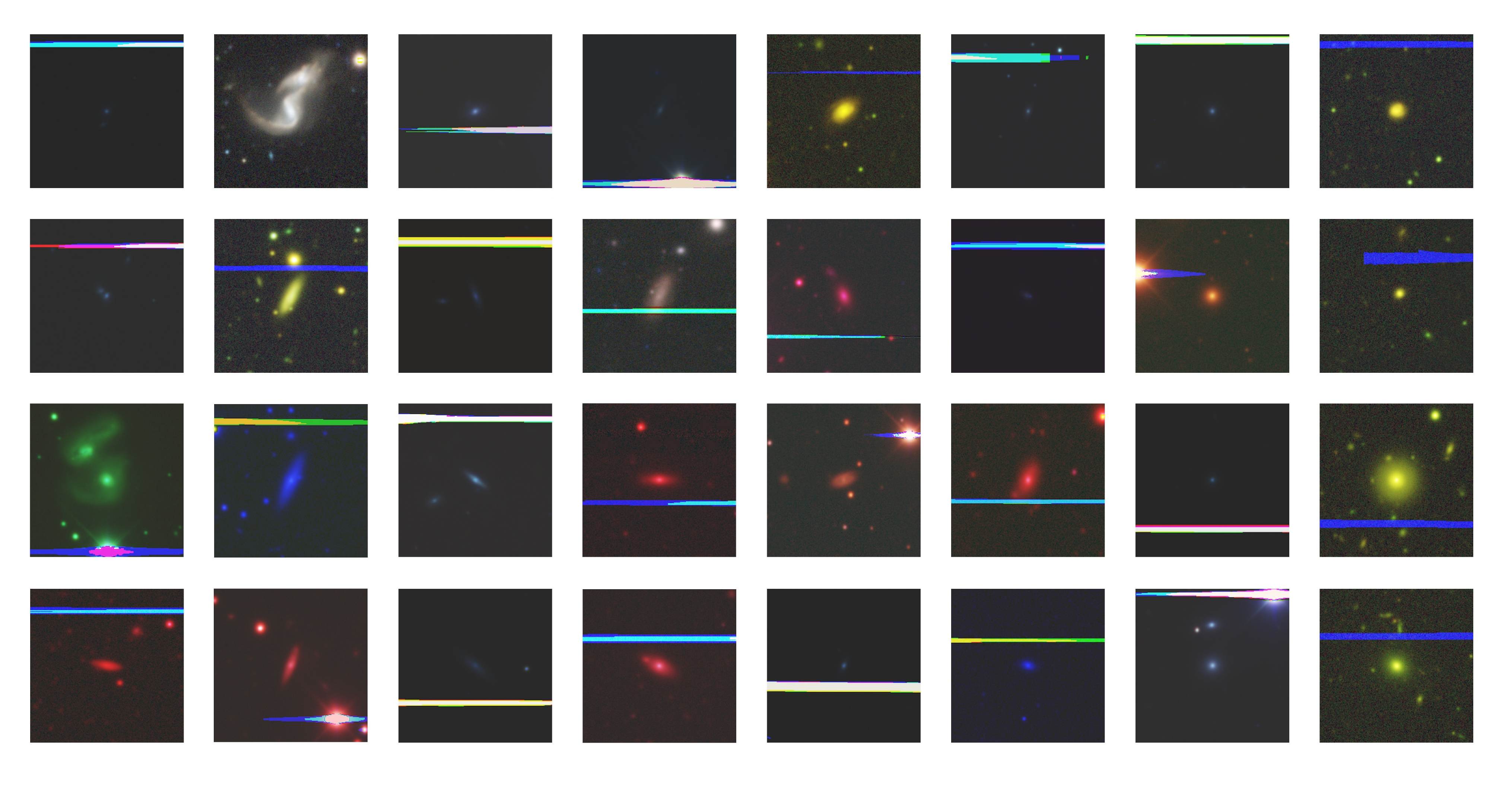}
    \caption{Examples of outlier galaxy images.}
    \label{fig:outlier}
\end{figure}

A consequence of creating dense distributions in the latent space is the ability to search for outliers located far from its centers. As an illustration, we identified outliers in the latent space using a criterion of eight times the standard deviation for each latent parameter. This yielded 1,308 outliers, representing 0.417\% of the total galaxy images. A random selection of these outliers is presented in Fig. \ref{fig:outlier}. We display 32 example images of outliers from the 40-dimensional latent variable space, chosen based on the $8$-$\sigma$ truncation criterion. Notably, the majority of these outliers are image defects, with the exception of the second image from the left in the top row, which depicts a merging galaxy affected by a bright star.

\subsection{Morphology Classification on Latent Variables}
\label{sec:classification}

The variational framework of the VAE model ensures that galaxy images with similar morphological features exhibit similar behavior in the latent space, leading to a dense distribution of latent variable features in this low-dimensional space.
Closely distributed and dense latent variables are not suitable for unsupervised clustering \citep[e.g.][]{RN10,nishikawa2020semi,wei2022unsupervised}. Therefore, in this section, we combine morphological labels to explore the morphological information contained in latent variables. 

Since latent variables already contain comprehensive information of galaxy images, we expect that directly using latent variables for morphological classification may simplify the classification logic and the number of operations. To test this idea, we adopt the random forest \citep{breiman2001random}, a classifier that uses multiple decision trees to train and predict samples, to classify the morphological labels using the latent variables. We employed the random forest algorithm from the \texttt{sklearn} package in Python, setting the number of decision trees to 100. Additionally, we adjusted the proportions of various categories to address the class imbalance. Specifically, we combined latent variables and morphological categories to form a dataset. We allocated 70\% of these data for training and the remaining 30\% for testing. During both training and testing phases, the random forests straightforwardly classified the latent variable features. To evaluate the effectiveness of the classification, we utilized precision, recall, and F-score as our performance metrics.

\begin{table}
	\centering
	\caption{Nine Random forests classification result of GZD-5 morphological feature questions.}
	\label{tab:rf_acc}
    \setlength{\tabcolsep}{1.5mm}{
	\begin{tabular}{cccccc}
		\hline
        \multirow{2}*{Question} & Number of & \multirow{2}*{Test acc} & \multirow{2}*{Precision} & \multirow{2}*{Recall} & \multirow{2}*{F1} \\
         & Features &  &  &  &  \\
		\hline
        How round & 3 & 83.69 \% & 83.69 \% & 83.69 \% & 83.69 \% \\
        Edge-on & 2 & 83.98 \% & 91.10 \% & 84.07 \% & 87.45 \% \\
        Bulge Shape & 3 & 76.94 \% & 76.94 \% & 76.94 \% & 76.94 \% \\
        Bar & 3 & 71.57 \% & 71.57 \% & 71.57 \% & 71.57 \% \\
        Have Arm & 2 & 88.51 \% & 93.47 \% & 89.03 \% & 91.19 \% \\
        Arm Tightness & 3 & 76.98 \% & 76.98 \% & 76.98 \% & 76.98 \% \\
        Arm Count & 3 & 71.28 \% & 71.28 \% & 71.28 \% & 71.28 \% \\
        Bulge Size & 3 & 80.82 \% & 80.82 \% & 80.82 \% & 80.82 \% \\
        Merger & 2 & 93.43 \% & 99.67 \% & 93.88 \% & 96.42 \% \\
        \hline
	\end{tabular}}
\end{table}

We train and test the morphological labels for nine morphological features (listed in Table \ref{tab:morph}) respectively. The accuracy of training and testing on the latent variables obtained by VAE is shown in Table \ref{tab:rf_acc}. In general, we obtain a test accuracy of between 70\% and 95\% for different classifications of morphological features. Compared with the deep learning algorithms directly applied to the morphologies of galaxy images with typical supervise accuracy \citep[\textgreater$90\%$, ][]{RN10}, the accuracy of our test on latent variables are slightly lower. However, most of the previous studies only make general morphology classifications for few global labels \cite[e.g. ][classification disk/elliptical/merging]{10.1093/mnras/stab1552}. Few studies have addressed the classification of various detail morphological features, and the resulting classification accuracy is also comparable \citep[e.g. active learning by][]{Walmsley2022} to us. Specifically, the accuracy of our simple classifier on the spiral arms' tightness, bulge size and roundness is similar to that of \citet{Walmsley2022} study. In contrast, the classification of the merging (93.43\% respectively) is even better.
 
Based on the above results, we see that these 40 latent variables output from the VAE dimensionality reduction algorithm can effectively retain the overall features of galaxy images (e.g. spiral/merging). 
Moreover, these features can be identified by simple classifiers such as random forests and so that used in the application of galaxy morphology classification. However, for these locally detailed morphological features (e.g. the arm tightness), they may not be effectively contained in low dimensional latent variables.

\section{Domain adaption: MGS in BASS and MzLS}\label{sec:domian_adapt}

Because of different observational conditions, the images of galaxies in the different surveys would be different in detail (e.g. image resolution, PSF, and noise level). Therefore, an optimized image encoder should try to extract the physical (morphological) features of the images while ignoring the fluctuations caused by observational conditions. Only if so, we expect that such an image encoder could be unbiasedly applied to galaxy images from other surveys. In this section, we test the generalization ability of the image encoder in the VAE model we have obtained in Section ~\ref{sec:application}, which is trained only by the DECaLS images and applies to BASS+MzLS images.

\subsection{Original encoder applied to BASS+MzLS galaxy images}
\label{sec:original}

The parameters for three different programs of the DESI image survey are listed in Table \ref{tab:desi}. We would like to remind you that the DECaLS survey covers the $g$, $r$, and $z$ bands, and all these bands have a uniform pixel resolution of $0\farcs262$. The BASS and MzLS surveys have the same footprint, but it differs from that of the DECaLS survey. The BASS survey operates in the $g$ and $r$ bands and has a pixel resolution of $0\farcs454$, whereas the MzLS survey is in the $z$ band with a pixel resolution of $0\farcs262$. The DECaLS survey and the combined BASS+MzLS survey overlap in a region of 2 degrees of declination, and the area of the overlapping sky covers about 100 square degrees.

In order to determine whether the encoder of the VAE algorithm trained on DECaLS images can be applied to BASS+MzLS $(g,r,z)$ images, we used the galaxies located within the overlapping footprint of the two surveys. Specifically, we selected 3,434 galaxies in the MGS ($r$\textless$17.77$) that were located within this overlapping region. We obtained images of these galaxies in both the MzLS and BASS surveys from the DESI website. To ensure consistency, we download the official offered $g$ and $r$ band BASS images with the same resolution as DECaLS, which was $0\farcs262$.

To compare the latent variables of these 3,434 galaxies in the two surveys, we applied the encoder trained in Section ~\ref{sec:dimension} to their $g,r,z$ images from BASS+MzLS and obtained their 40-dimensional latent variables. We then compared this new set of latent variables with those obtained from DECaLS images. To quantify the differences between the latent variables for each galaxy, we used Mean Squared Error (MSE).

\begin{figure*}
    \centering
    \subfigure[Similar latent variables (MSE \textless 0.1).]{
        \includegraphics[width=0.975\columnwidth]{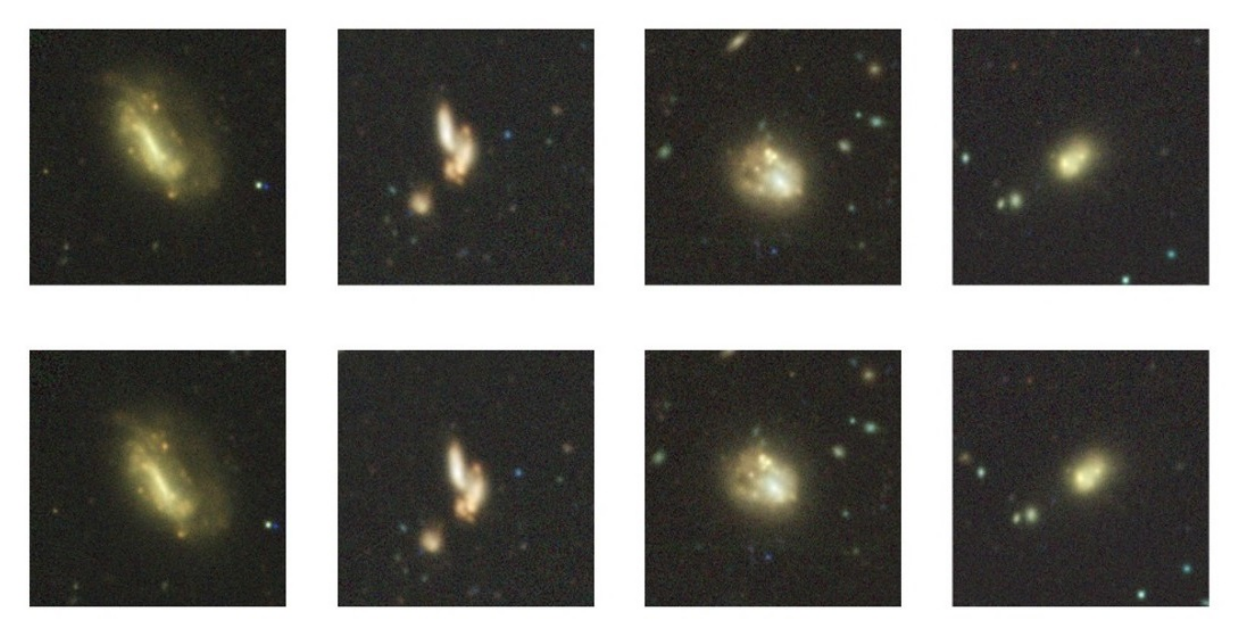}
        \label{fig:latent_like}
    }
    \subfigure[Dissimilar Latent variables (MSE \textgreater 8).]{
        \includegraphics[width=0.975\columnwidth]{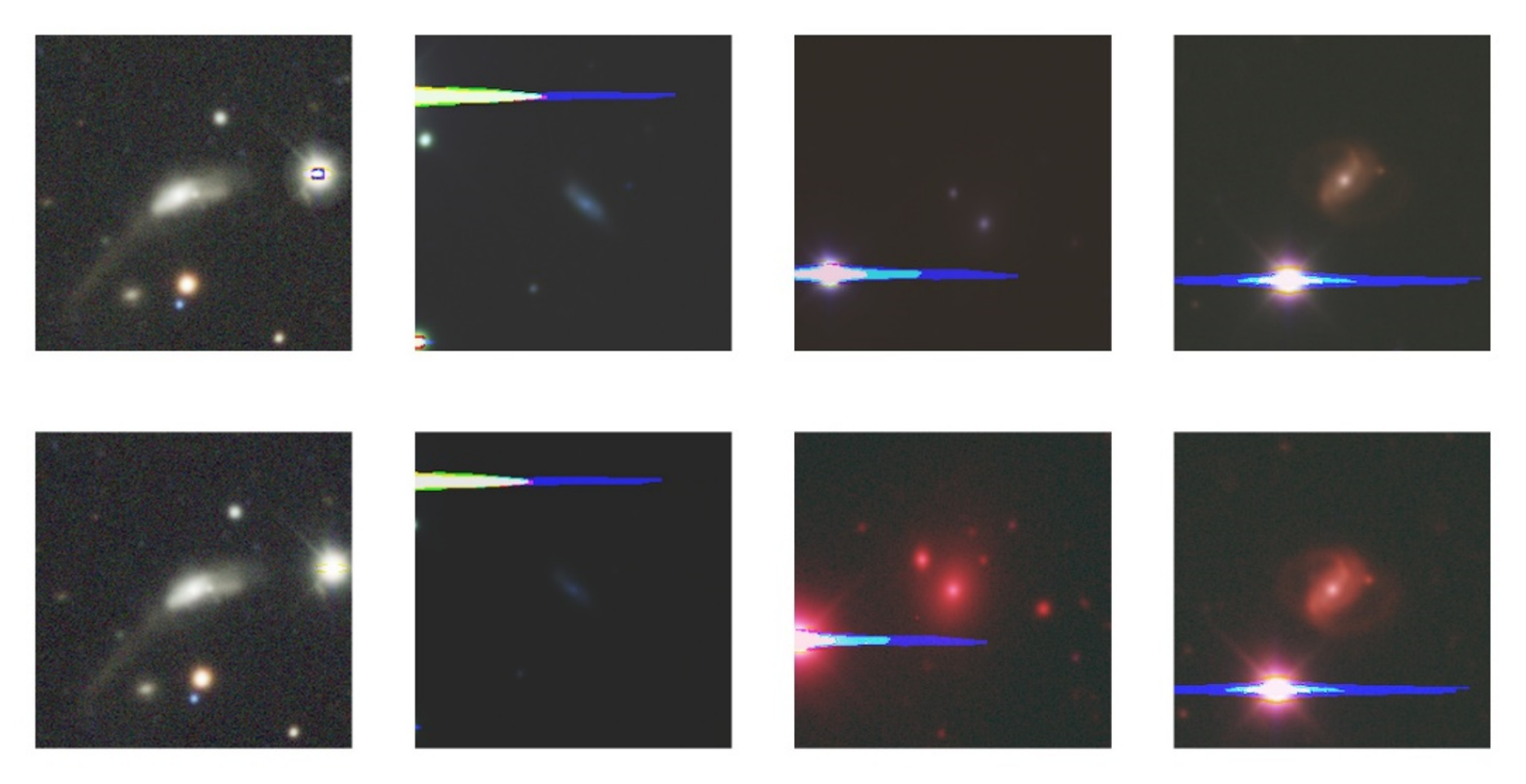}
        \label{fig:latent_dislike}
    }%
    \caption{Image comparison of the same galaxies observed by DECaLS (top panels) and BASS+MzLS (bottom panels). (a) Galaxies with similar latent variables, i.e. MSE \textless 0.1; (b) Galaxies with dissimilar latent variables, i.e. MSE \textgreater 8.}
    \label{fig:differ}
\end{figure*}

Although the images of these galaxies appear almost identical in the two different surveys (DECaLS vs. BASS+MzLS), we found differences in their latent variables. The MSE values of the latent variables ranged from as small as $\sim0.05$ to as large as 10. We show two groups of images with very different MSE values in Figure \ref{fig:differ}: one group with MSE \textless$0.1$ and another group with MSE \textgreater$8$. As expected, the group with small MSE values had almost identical images and showed little visual difference. For the group with very large MSE values, we observed contamination by a bright star nearby in most cases. We suspect that it is the bright star that led to significant differences in the latent variables after dimensionality reduction by the VAE encoder.

We further tested whether these latent variables of BASS+MzLS galaxies can be used to make morphology classifications, as we did in Section ~\ref{sec:classification}. We used the same random forest that was trained in Section ~\ref{sec:classification} to test the morphology classification of these 3,434 galaxies. For simplicity, we only applied the random forest on some morphology features where the number of galaxy samples was larger than 100. The results are shown in Table \ref{tab:vae_domain}.

As expected, compared to the DECaLS galaxy images, the morphological classification of the latent variables of BASS+MzLS galaxy images showed a significant decrease. In particular, for bar features of galaxies, the classification precision dropped from 82.53\% to 70.59\%, while for the merger features, the classification accuracy remained almost unchanged. This result suggests that the VAE encoder has successfully extracted some of the global morphology features of galaxy images (e.g., mergers). In contrast, random noises generated during observations affect some detailed morphology features (e.g., bar).

\subsection{Optimized Encoder with Domain Adaptation}
\label{sec:domainadapt}

In the above section, we have demonstrated that slight differences in survey parameters prevent us from using the VAE encoder trained by DECaLS only on BASS+MzLS galaxies, even though the target galaxies are a homogeneous sample in both surveys.
To solve this problem, we introduce the idea of domain adaption and MMD algorithm (Section ~\ref{sec:DA}). In our problem, we have a well-trained network in the source domain (DECaLS), which can well implement image feature extraction, i.e. dimensionality reduction process. What we need is a fine-tuning or migration of the network to make it suitable for the dimensionality reduction of galaxy images in both the source domain (DECaLS) and the target domain (BASS+MzLS).

\begin{figure}
	\includegraphics[width=\columnwidth]{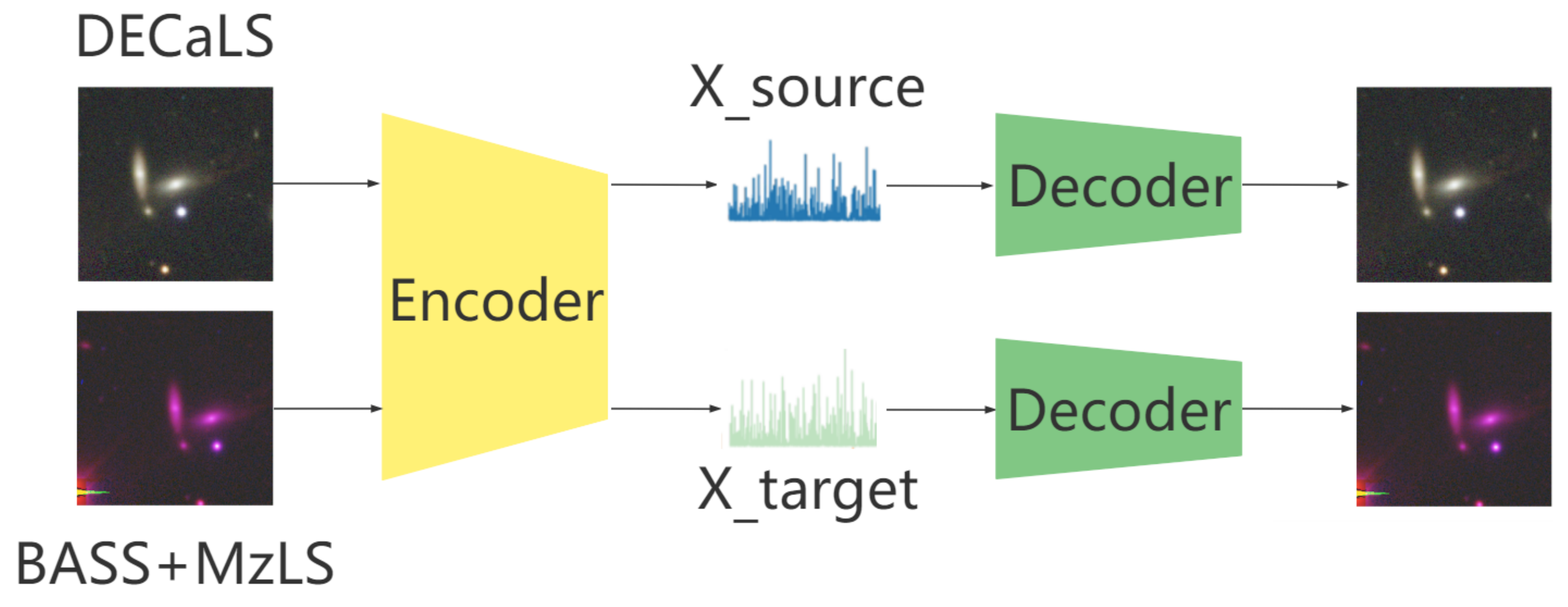}
    \caption{Structural details of the Domain Adaptation based on VAE in this study.}
    \label{fig:vae_domian}
\end{figure}

In our model, this difference is the difference of latent variables obtained by the same VAE encoder for the same galaxies in two groups of surveys. 
By combining MMD with the existing VAE network (Fig. \ref{fig:vae}), we obtain a domain adaptation model, which is illustrated in Figure \ref{fig:vae_domian}. In this network, the latent variables of the galaxy images in two different surveys are first extracted by the same encoder. Then two decoders are attached to reconstruct the images of the galaxies themselves. Using such a domain adaption model, the latent variables obtained by the encoder will better reflect the internal morphological features of the galaxy itself. In contrast, the differences in the galaxy images caused by different survey parameters are contained in two other decoders.

To perform domain adaptation, we augment the original VAE loss function with Maximum Mean Discrepancy (MMD) as follows:
\begin{equation}\label{transfer}
\mathcal{L}_{DA} = \mathcal{L}_{VAE} + \lambda \cdot \mathrm{MMD}^2(p_\phi, X_S, X_T)\,
\end{equation}
where $\lambda$ is the super-parameter that balanaces the loss from VAE and MMD networks,  $p_\phi$ denotes the encoder mapping that we aim to optimize.  This domain-adapted deep learning network is based on the weights of the VAE model trained on the DECaLS (Section ~\ref{sec:application}) and trained on the images of 3,434 galaxies in the overlapping regions of DECaLS and BASS+MzLS, which aims to learn latent variables (image features) that are invariant to differences between two surveys, allowing us to compare and analyze the same galaxies in different surveys \citep[e.g.][]{pan2010domain}.  Here, we set $\lambda=10$ so that the final loss of MMD is of the same magnitude as that of the VAE network. For the training, we also set the learning rate to 0.0001 and used a smaller number of epoch 20.

\begin{table}
	\centering
	\caption{Comparison of random forest classification accuracy before and after VAE domain adaptation.}
	\label{tab:vae_domain}
    \setlength{\tabcolsep}{0.8mm}{
	\begin{tabular}{cc|cc|cc}
		\hline
        \multirow{2}*{Question} & Total & Before DA & After DA & Before DA & After DA \\
         & Count & DECaLS & DECaLS & BASS+MzLS & BASS+MzLS \\
        \hline
        How round & 625 & 76.59\% & \textbf{82.45 \%} & 74.47 \% & \textbf{77.66 \%}  \\
        Edge-on & 1024 & 88.60 \% & \textbf{90.23 \%} & 87.62 \% & \textbf{89.25 \%}  \\
        Bar & 503 & \textbf{96.69 \%} & 96.03 \% & 92.72 \% & \textbf{94.70 \%} \\
        Have Arm & 784 & \textbf{83.40 \%} & 82.55 \% & 80.43 \% & \textbf{84.68 \%} \\
        Arm Tightness & 225 & 61.76 \% & \textbf{67.65 \%} & 66.18 \% & \textbf{69.12 \%} \\
        Bulge Size & 285 & 73.26 \% & \textbf{76.74 \%} & 70.93 \% & \textbf{74.42 \%} \\
        Merger & 1135 & 97.84\% & \textbf{97.95 \%} & 97.65 \% & \textbf{98.53 \%} \\
        \hline
	\end{tabular}}
\end{table}

To evaluate and compare the effectiveness of domain adaptation, we use the same random forest algorithm as in Section ~\ref{sec:classification} to test the performance of morphological classification based on the learned latent variables. Since the latent variables are obtained from the newly adapted network, we retrain the random forest with 70\% of the data for training and 30\% for testing. This approach allows us to assess whether the domain adaptation process leads to more effective representations of galaxy images for classification purposes.

Given that this domain-adapted network was trained on only 3,434 galaxies, the number of galaxies with certain morphological features is quite limited. This restricts the effective training of subsequent random forest classification algorithms. For effective training, we required a minimum of $\gtrapprox$ 200 galaxies. Consequently, only 7 out of the 9 galaxy features meet this criterion, as shown in Table \ref{tab:vae_domain}.

We utilize the accuracy of the random forest to assess the feature extraction capability of the encoder. After implementing domain adaptation on the VAE model, there was a notable enhancement in the precision of the morphological features for the BASS+MzLS data, as depicted in Table \ref{tab:vae_domain}. For the DECaLS data, post-domain adaptation, precision improved for 5 out of the 7 features.
Interestingly, when we compare the results of DECaLS data before domain adaptation to those of BASS+MzLS data after domain adaptation, we notice enhancements in the precision metrics for most features. Specifically, the precision for the 'arm tightness' feature rose from 61.76\% to 69.12\%, and the precision for the 'merger' feature went up from 97.95\% to 98.53\%.

These results suggest that domain adaptation enables deep learning networks to extract more relevant internal features from galaxy images while reducing the influence of random fluctuations introduced by different observations. The increased classification accuracy provides evidence to support this idea.

\section{Conclusion and discussion}
\label{sec:conslusion}

In this study, we applied VAE to the $g,r,z$ images of galaxies from the DECaLS survey, which were classified based on visual inspection. We found that a VAE algorithm with a latent variable dimension of $N_L=40$ was effective in extracting morphological features from galaxy images. 

To evaluate the quality of the reconstructed images, we used the structural similarity index (SSIM) and found that the VAE reconstructed images had SSIM values between 0.80-0.90 for all morphological types of galaxies, indicating good image reconstruction.  

After reducing the dimensionality of the galaxy images to a 40-dimensional latent variable space, we used a classical random forest classifier to make detailed morphological classifications. The test accuracy varied between 0.71 to 0.94, depending on the morphological types. In general, the latent variables performed well on global morphological features (e.g. round, merger) and were slightly weaker on detailed features (e.g. bar/spiral arm).

To apply our VAE encoders to both the DECaLS and BASS+MzLS surveys, we tuned our network and adapted it for both surveys by adding MMD of the latent variables as a loss component. We then trained our algorithm using 3,434 galaxies in the overlapping footprint of the surveys. The results showed that our VAE algorithm was successfully adapted to the galaxy images of both surveys.

The latent variables extracted by the VAE encoder after domain adaptation (DA) performed better on galaxy morphology classification, indicating that the VAE encoder after DA was better at extracting internal features of galaxy morphology and was less affected by observational noise.

Our VAE framework has many potential applications in observational studies of galaxy morphologies.

First, the prior distribution of latent variables in the VAE algorithm ensures that galaxy images with similar latent variables are highly similar. Therefore, we can easily identify outliers in the galaxy images using the latent variables, such as images with defects or peculiar morphologies.
As shown in Fig. \ref{fig:outlier}, most of the outliers were images with defects. This demonstrates the ability of the VAE dimensionality reduction method to quickly and effectively detect defected images in modern and future image surveys, such as the CSST optical survey \citep{Sun_2021}. In addition, we can also use this method to identify galaxies with peculiar morphologies, which will be the subject of a future study.

Secondly, our VAE decoders can be utilized to generate mock images at low redshifts. Similar studies have been conducted using deep galaxy images from the HST/ACS COSMOS survey \citep{lanusse2021deep}. However, generating mock images that accurately reflect the characteristics of observed galaxies is challenging due to the varying redshift and apparent size distributions of galaxies in image surveys with different depths. Therefore, a VAE generative model based on shallow-ground survey data can provide a complementary approach to that of \citet{lanusse2021deep}.

In this study, we have demonstrated that the VAE model with DA can successfully be applied to both DECaLS and BASS+BzLS galaxy images. However, this finding alone does not justify the direct application of our VAE encoders to new sky surveys.
The reason is that different surveys may have significantly different PSFs which moreover are expected to be position and time-dependent, as the instrument and site conditions evolve. The real, intrinsic image signals of astronomical objects are convolved with these highly complex PSFs, and this complexity is not expected to be easily captured by our current network. Additionally, different surveys may have varying depths, meaning that fainter galaxies and those at higher redshifts may emerge as the survey depth increases. Thus, training an encoder of galaxy images that can be adapted to all galaxies in entirely different sky surveys requires further investigation. Nevertheless, the results presented here are evidence that the coherent combined application of machine learning concepts (VAE and DA) may be one possible path for the future development of multi-survey analysis.

\section*{Acknowledgements}

This work is supported by research grants from the China Manned Space Project with NO. CMS-CSST-2021-A07 and program of Shanghai Academic/Technology Research Leader (22XD1404200). We also acknowledge the grants from the National Natural Science Foundation of China (No. 12073059 $\&$ No. U2031139), the National Key R$\&$D Program of China (No. 2019YFA0405501, 2022YFF0503402). RSS acknowledges the grants from the Chinese Academy of Sciences Hundred Talents Project (E12501100C). We thank the referee, for useful comments that have helped improve the paper.

The Legacy Surveys consist of three individual and complementary projects: the Dark Energy Camera Legacy Survey (DECaLS; Proposal ID \#2014B-0404; PIs: David Schlegel and Arjun Dey), the Beijing-Arizona Sky Survey (BASS; NOAO Prop. ID \#2015A-0801; PIs: Zhou Xu and Xiaohui Fan), and the Mayall z-band Legacy Survey (MzLS; Prop. ID \#2016A-0453; PI: Arjun Dey). DECaLS, BASS and MzLS together include data obtained, respectively, at the Blanco telescope, Cerro Tololo Inter-American Observatory, NSF’s NOIRLab; the Bok telescope, Steward Observatory, University of Arizona; and the Mayall telescope, Kitt Peak National Observatory, NOIRLab. Pipeline processing and analyses of the data were supported by NOIRLab and the Lawrence Berkeley National Laboratory (LBNL). The Legacy Surveys project is honored to be permitted to conduct astronomical research on Iolkam Du’ag (Kitt Peak), a mountain with particular significance to the Tohono O’odham Nation.

NOIRLab is operated by the Association of Universities for Research in Astronomy (AURA) under a cooperative agreement with the National Science Foundation. LBNL is managed by the Regents of the University of California under contract to the U.S. Department of Energy.

This project used data obtained with the Dark Energy Camera (DECam), which was constructed by the Dark Energy Survey (DES) collaboration. Funding for the DES Projects has been provided by the U.S. Department of Energy, the U.S. National Science Foundation, the Ministry of Science and Education of Spain, the Science and Technology Facilities Council of the United Kingdom, the Higher Education Funding Council for England, the National Center for Supercomputing Applications at the University of Illinois at Urbana-Champaign, the Kavli Institute of Cosmological Physics at the University of Chicago, Center for Cosmology and Astro-Particle Physics at the Ohio State University, the Mitchell Institute for Fundamental Physics and Astronomy at Texas A\&M University, Financiadora de Estudos e Projetos, Fundacao Carlos Chagas Filho de Amparo, Financiadora de Estudos e Projetos, Fundacao Carlos Chagas Filho de Amparo a Pesquisa do Estado do Rio de Janeiro, Conselho Nacional de Desenvolvimento Cientifico e Tecnologico and the Ministerio da Ciencia, Tecnologia e Inovacao, the Deutsche Forschungsgemeinschaft and the Collaborating Institutions in the Dark Energy Survey. The Collaborating Institutions are Argonne National Laboratory, the University of California at Santa Cruz, the University of Cambridge, Centro de Investigaciones Energeticas, Medioambientales y Tecnologicas-Madrid, the University of Chicago, University College London, the DES-Brazil Consortium, the University of Edinburgh, the Eidgenossische Technische Hochschule (ETH) Zurich, Fermi National Accelerator Laboratory, the University of Illinois at Urbana-Champaign, the Institut de Ciencies de l’Espai (IEEC/CSIC), the Institut de Fisica d’Altes Energies, Lawrence Berkeley National Laboratory, the Ludwig Maximilians Universitat Munchen and the associated Excellence Cluster Universe, the University of Michigan, NSF’s NOIRLab, the University of Nottingham, the Ohio State University, the University of Pennsylvania, the University of Portsmouth, SLAC National Accelerator Laboratory, Stanford University, the University of Sussex, and Texas A\&M University.

BASS is a key project of the Telescope Access Program (TAP), which has been funded by the National Astronomical Observatories of China, the Chinese Academy of Sciences (the Strategic Priority Research Program “The Emergence of Cosmological Structures” Grant \# XDB09000000), and the Special Fund for Astronomy from the Ministry of Finance. The BASS is also supported by the External Cooperation Program of Chinese Academy of Sciences (Grant \# 114A11KYSB20160057), and Chinese National Natural Science Foundation (Grant \# 12120101003, \# 11433005).

The Legacy Survey team makes use of data products from the Near-Earth Object Wide-field Infrared Survey Explorer (NEOWISE), which is a project of the Jet Propulsion Laboratory/California Institute of Technology. NEOWISE is funded by the National Aeronautics and Space Administration.

The Legacy Surveys imaging of the DESI footprint is supported by the Director, Office of Science, Office of High Energy Physics of the U.S. Department of Energy under Contract No. DE-AC02-05CH1123, by the National Energy Research Scientific Computing Center, a DOE Office of Science User Facility under the same contract; and by the U.S. National Science Foundation, Division of Astronomical Sciences under Contract No. AST-0950945 to NOAO.

\section*{Data Availability}
 
This work uses morphological labels from the fraction thresholds, which are released by \cite{Walmsley2022} on \url{https://doi.org/10.5281/zenodo.4573248}. The galaxy image (DECaLS, BASS and MzLS) data underlying this article are available at \url{https://www.legacysurvey.org/}. Our code is publicly available: \url{https://github.com/xuquanfeng/Galaxy_VAE}.










\bsp	
\label{lastpage}

\begin{thebibliography}{50}
\providecommand\natexlab[1]{#1}
\providecommand\JournalTitle[1]{#1}

\bibitem[Searle {et~al.}(1973)]{1973ApJ...179..427S} Searle L., Sargent W.~L.~W., Bagnuolo W.~G., 1973,
  \apj, 179, 427

\bibitem[Dressler (1980)]{1980ApJ...236..351D} Dressler A., 1980,
  \apj, 236, 351

\bibitem[Strateva {et~al.}(2001)]{Strateva_2001} Strateva I., Ivezi{\'{c}} {\v{Z}}., {et~al.} 2001,
  \apj, 122, 1861

\bibitem[Petrillo {et~al.}(2017)]{10.1093/mnras/stx2052} Petrillo C. E., Tortora C., Chatterjee S., 2017,
  \mnras, 472, 1129

\bibitem[Aniyan \& Thorat(2017)]{Aniyan_2017}Aniyan A. K., Thorat K., 2017, 
  The Astrophysical Journal Supplement Series, 230, 20

\bibitem[{\'{C}}iprijanovi{\'{c}} {et~al.}(2020)]{CIPRIJANOVIC2020100390}{\'{C}}iprijanovi{\'{c}} A., Snyder G.F., Nord B., {et~al.} 2020,
  The Astrophysical Journal, 32, 100390

\bibitem[{\'{C}}iprijanovi{\'{c}} {et~al.}(2023)]{ciprijanovic2023deepastrouda}{\'{C}}iprijanovi{\'{c}} A., Lewis A., Pedro K., {et~al.} 2023,
  Machine Learning: Science and Technology, 4, 025013

\bibitem[Vilalta {et~al.}(2019)]{Vilalta_2019}Vilalta R., Gupta K.D., Boumber D., {et~al.} 2019,
  Publications of the Astronomical Society of the Pacific, 131, 108008

\bibitem[{\'{C}}iprijanovi{\'{c}} {et~al.}(2021)]{10.1093/mnras/stab1677}{\'{C}}iprijanovi{\'{c}} A., 2019,
  Monthly Notices of the Royal Astronomical Society, 506, 677

\bibitem[Ganin {et~al.}(2016)]{ganin2016domain}Ganin Y., Ustinova E., Ajakan H., {et~al.} 2016,
  The journal of machine learning research, 17, 2096

\bibitem[Gilda {et~al.}(2022)]{gilda2022unsupervised}Gilda S., de Mathelin A., Bellstedt S., {et~al.} 2022,
  Unsupervised Domain Adaptation for Constraining Star Formation Histories

\bibitem[Zhu {et~al.}(2019)]{RN10}Zhu X.P., Dai J.M.,Bian C.J., {et~al.} 2019,
  Astrophysics and Space Science, 364, 55

\bibitem[Willett {et~al.}(2013a)]{Willett2013}Willett K.W., Lintott C.J., Bamford S.P., 2013,
  \mnras, 435, 2835

\bibitem[Dey {et~al.}(2019)]{Dey_2019}Dey A., Schlegel D. J., Lang D., {et~al.} 2019,
  The Astronomical Journal, 157, 168

\bibitem[Cavanagh \& Bekki{et~al.}(2020)]{cavanagh_bars_2020}Cavanagh M.K., Bekki K.,2020,
  Astronomy \& Astrophysics, 641, A77

\bibitem[Mr{\'{o}}z(2020)]{mroz_identifying_2020}Mr{\'{o}}z P., 2020,
  Acta Astronomica, 70, 169

\bibitem[Vavilova {et~al.}(2021)]{vavilova2021machine}Vavilova I., Dobrycheva D., Vasylenko M.Y., {et~al.} 2021,
  Astronomy \& Astrophysics, 648, A122

\bibitem[Ahn {et~al.}(2013)]{Ahn2012}Ahn C.P., Alexandroff R., Allende P.C., 2012,
  \apjs, 203, 21

\bibitem[Huertas-Company {et~al.}(2018)]{Huertas-Company_2018}Huertas-Company M., Primack J.R., Dekel A., {et~al.} 2018,
  The Astrophysical Journal, 858, 114

\bibitem[Cheng {et~al.}(2021b)]{10.1093/mnras/stab2142}Cheng T.Y., Conselice C.J.,Arag{\'{o}}n-Salamanca A., {et~al.} 2021,
  \mnras, 507, 4425

\bibitem[Li(2022a)]{10.1093/mnras/stac2697}Li J., Tu L.P., Gao X., {et~al.} 2022,
  \mnras, 517, 808

\bibitem[Ghosh(2021)]{2022ApJ...935..138G}Ghosh A., Urry C.M., Rau A., {et~al.} 2022,
  \apj, 935, 138

\bibitem[Li(2022b)]{Li_2022}Li R., Napolitano N. R., Roy N., {et~al.} 2022,
  \mnras, 517, 808

\bibitem[Boucaud(2020)]{Boucaud2020}Boucaud A., Huertas-Company M., Heneka C., 2020,
  \mnras, 491, 2481

\bibitem[Wang {et~al.}(2022)]{2022PhRvD.106f3023W}Wang H., Sreejith S., Slosar A., {et~al.} 2022,
  Physical Review D, 106, 063023

\bibitem[Martin {et~al.}(2019)]{10.1093/mnras/stz3006}Martin G., Kaviraj S., Hocking A., 2019,
  \mnras, 491, 1408

\bibitem[Hayat {et~al.}(2021)]{hayat2021self}Hayat M.A., Stein G., Harrington P., 2021,
  The Astrophysical Journal Letters, 911, L33

\bibitem[Wei {et~al.}(2022)]{wei2022unsupervised}Wei S.L., Li Y.D., Lu W., {et~al.} 2022,
  Publications of the Astronomical Society of the Pacific, 134, 114508

\bibitem[Fielding {et~al.}(2022)]{fielding2022classification}Fielding E., Nyirenda C.N., Vaccari M., 2022,
  in 2022 International Conference on Electrical, Computer and Energy Technologies (ICECET), pp, 1-6

\bibitem[Hocking {et~al.}(2018)]{2018MNRAS.473.1108H}Hocking A., Geach J. E., Sun Y., {et~al.} 2018,
  \mnras, 473, 1108

\bibitem[Hovis-Afflerbach {et~al.}(2021)]{Hovis_Afflerbach_2021}Hovis-Afflerbach B., Steinhardt C.L., Masters D., {et~al.} 2021,
  The Astrophysical Journal, 908, 148

\bibitem[Kartaltepe {et~al.}(2015)]{Kartaltepe_2015}Kartaltepe J.S., Mozena M., Kocevski D., {et~al.} 2015,
  The Astrophysical Journal Supplement Series, 221, 11

\bibitem[Uzeirbegovic {et~al.}(2020)]{10.1093/mnras/staa2651}Uzeirbegovic E., Geach J.E., Kaviraj S., {et~al.} 2020,
  \mnras, 498, 4021

\bibitem[Simmons {et~al.}(2016)]{Simmons2016}Simmons B.D., Lintott C., Willett K.W., {et~al.} 2016,
  \mnras, 464, 4420

\bibitem[Kingma \& Welling {et~al.}(2013)]{kingma2013auto}Kingma D.P., Welling M., 2013,
  arXiv preprint arXiv:1312.6114

\bibitem[Lanusse {et~al.}(2021)]{lanusse2021deep}Lanusse F., Mandelbaum R., Ravanbakhsh S., {et~al.} 2021,
  \mnras, 504, 5543

\bibitem[Cheng {et~al.}(2021a)]{10.1093/mnras/stab734}Cheng T.Y., Huertas-Company M., Conselice C.J., {et~al.} 2021,
  \mnras, 503, 4446

\bibitem[Zhou {et~al.}(2021a)]{Zhou_2022}Zhou C.C., Gu Y.M., Fang G.W., {et~al.} 2022,
  The American Astronomical Society, 163, 86

\bibitem[{\'C}iprijanovi{\'c} {et~al.}(2022)]{ciprijanovic2022deepadversaries}{\'C}iprijanovi{\'c} A., Kafkes D., Snyder G., {et~al.} 2022,
  Machine Learning: Science and Technology, 3, 035007

\bibitem[{\'C}iprijanovi{\'c} {et~al.}(2023)]{arXiv:2302.02005}{\'C}iprijanovi{\'c} A., Lewis A., Pedro K., {et~al.} 2023,
  arXiv preprint arXiv:2302.02005

\bibitem[Strauss {et~al.}(2002)]{Strauss_2002}Strauss M.A., Weinberg D.H., Lupton R.H., {et~al.} 2002,
  The Astrophysical Journal,  124, 1810

\bibitem[Abazajian {et~al.}(2009)]{Abazajian_2009}Abazajian K.N., Adelman-McCarthy J.K., Agüeros M.A., {et~al.} 2009,
  The Astrophysical Journal Supplement Series, 182, 543

\bibitem[Willett {et~al.}(2013b)]{10.1093/mnras/stt1458}Willett K.W., Lintott C.J., Bamford S.P., {et~al.} 2013,
  \mnras, 435, 2835

\bibitem[Kullback \& Leibler {et~al.}(1951)]{Kullback1951}Kullback S., Leibler R.A., {et~al.} 1951,
  Ann. Math. Statist., 22, 79

\bibitem[Pan {et~al.}(2008)]{pan2008transfer}Pan S.J., Kwok J.T., Yang Q., {et~al.} 2008,
  in AAAI, pp, 677–682

\bibitem[Wang {et~al.}(2004)]{wang2004image}Wang Z., Bovik A.C., Sheikh H.R., {et~al.} 2004,
  IEEE transactions on image processing, 13, 600

\bibitem[Nishikawa-Toomey {et~al.}(2020)]{nishikawa2020semi}Nishikawa-Toomey M., Smith L., Gal Y., {et~al.} 2020,
  arXiv preprint arXiv:2011.08714

\bibitem[Breiman {et~al.}(2001)]{breiman2001random}Breiman L., {et~al.} 2001,
  Machine learning, 45, 5

\bibitem[Cavanagh {et~al.}(2021)]{10.1093/mnras/stab1552}Cavanagh M.K., Bekki K., Groves B.A., {et~al.} 2021,
  \mnras, 506, 659

\bibitem[Walmsley {et~al.}(2022)]{Walmsley2022}Walmsley M., Lintott C., G{\'{e}}ron T., {et~al.} 2021,
  \mnras, 509, 3966

\bibitem[Pan {et~al.}(2010)]{pan2010domain}Pan S.J., Tsang I.W., Kwok J.T., {et~al.} 2010,
  IEEE transactions on neural networks, 22, 199

\bibitem[Sun {et~al.}(2021)]{Sun_2021}Sun Y., Deng D.S., Yuan H.B., {et~al.} 2021,
  National Astronomical Observatories, CAS and IOP Publishing Ltd., 21, 092



\end{thebibliography}
\end{document}